# MELANINA: UM PIGMENTO NATURAL MULTIFUNCIONAL

João V. Paulin,[a] Barbara Fornaciari,[b] Bruna A. Bregadiolli,[a] Mauricio S. Baptista,[b] Carlos F. O. Graeff,[a,*]

[a] Departamento de Física, Faculdade de Ciências, Universidade Estadual Paulista (UNESP), 17033-360 Bauru - SP, Brasil.

[b] Departamento de Bioquímica, Instituto de Química, Universidade de São Paulo (USP), 05508-900 São Paulo - SP, Brasil.

*e-mail:* carlos.graeff@unesp.br

# MELANIN: A MULTIFUNCTIONAL NATURAL PIGMENT

The melanins are a ubiquitous class of pigments found throughout nature from lower organisms to humans. It is one of the few biopolymers with fascinating functions in biology, medicine, chemistry, physics and engineering due to its relation to neurological diseases, skin disorders and melanoma, and, in the last couple of decades, its applications in biocompatible and biodegradable devices for organic electronic and bioelectronics. In this review, we present the major advances in physicochemical, biochemical and photochemical properties as well as some melanin-based technological application.

**INTRODUÇÃO**

Melanina é uma classe de biomacromolécula funcional derivada de quinonas e fenóis, que é encontrada por toda fauna (mamíferos, insetos, fungos e bactérias) e flora com um conjunto interessante de propriedades físico-químicas.[1,2] No corpo humano a melanina apresenta importantes funções como supressão de radicais livres, fotoproteção, pigmentação, quelação de íons metálicos, termoregulação e transmissão de sinais neurais.[1–4] Além disso, as melaninas parecem estar relacionada de alguma forma relacionadas ao melanoma, um tipo de câncer de pele com grandes chances de progredir para metástase, bem como com alguns tipos de distúrbios neurológicos.[3,5–8] Além de suas importantes funções biológicas, a melanina tem grande potencial tecnológico como nos materiais funcionais para aplicações em bioeletrônica.

Este pigmento não é recente na nossa escala evolutiva. Vários estudos paleontológicos mostram que as organelas responsáveis pela produção da melanina, os melanossomos, estavam presentes desde o final da Era Jurássica (145-161 milhões de anos atrás).[9–11] De fato, o melanossomo foi observado em um fóssil na região do Ceará no Brasil, de um pterodátilo *Tupandactylus imperator,* datado de 110 milhões de anos.[12]

O termo "melanina" (do grego antigo $\mu\acute{\epsilon}\lambda\alpha\varsigma$ cuja tradução é preto) foi utilizado pela primeira vez em 1840 por Berzelius, um químico sueco, para se referir aos pigmentos presentes em animais de coloração escura, e desde então é comumente utilizado para descrever todo tipo de pigmento natural preto ou marrom escuro, sem ter necessariamente uma relação clara com as suas características químicas ou biológicas.[13] Desta forma, com o passar dos anos, um conjunto de definições foram propostas para melhor classificar esta ampla classe de biomateriais.

O principal grupo de melaninas é conhecido como *eumelanina*. Eumelanina é um pigmento insolúvel de coloração preto-marrom, sintetizado a partir da tirosina ou da 3,4-dihidroxi-L-fenilalanina (L-DOPA), através de estruturas intermediárias do tipo 5,6-dihidroxi-indol. Um segundo grupo é chamado *feomelanina, sendo* sintetizado também a partir da tirosina ou L-DOPA, porém na presença de L-cisteína, dando origem a monômeros de benzotiazina e benzotiazol, unidades que, além do nitrogênio, contém enxofre em sua estrutura.[14,15] Este grupo apresenta uma coloração que varia entre vermelho escuro e amarelo, sendo encontrados predominantemente em indivíduos ruivos.[14] *Alomelanina*, isto é, "outra"-melanina, categoriza o terceiro grande grupo das melaninas. Alomelaninas são pigmentos escuros produzidos por insetos, microorganismos e plantas.[15–18] que não tem precursores sintéticos bem definidos, sendo obtidos a partir da oxidação/polimerização de catecóis, 2,5-dihidroxifenil-acetato (homogentisato) 1,8-dihidroxinaftaleno, γ-glutaminil-4-hidroxibenzeno e ácidos 4-hidroxifenilacético.

Por fim, como mencionado anteriormente, as melaninas também estão presentes no sistema nervoso central e nas glândulas suprarrenais.[15,19–21] Neste caso, a melanina é chamada neuromelanina e está presente em grande quantidade nos seres humanos, em menor quantidade em alguns primatas e totalmente ausente em diversas outras espécies.[4] Neuromelanina é considerada um subgrupo das melaninas, sendo uma partícula escura composta por feomelanina em seu interior e eumelanina na camada mais externa.[19–21] Nas últimas décadas, a neuromelanina passou de um simples composto inerte sem função definida, para um importante elemento da substância negra ligado a doenças neurodegenerativas, como a doença de Parkinson.[15,2223] Atualmente, sugere-se que a interação entre a neuromelanina e íons metálicos diminua o estresse oxidativo do meio, limitando, como consequência, a degeneração dos neurônios.[23]

Nesta revisão apresentaremos um apanhado das propriedades físicas, químicas e bioquímicas da melanina e dos seus derivados sintéticos, além de apresentar algumas de suas aplicações nos campos da eletrônica orgânica e bioeletrônica. Nosso foco será principalmente no grupo da eumelanina, de forma que os termos "eumelanina" e "melanina" serão utilizados indistintamente, a não ser quando definirmos se tratar de um tipo de específico de melanina.

**ESTRUTURA**

A estrutura da melanina natural ainda é objeto de discussão na literatura. Contudo, sabe-se que a melanina é composta por unidades de 5,6-dihidroxi-indol em diferentes estados redox, Figura 1. Essas unidades formam principalmente estruturas covalentemente ligadas (com até 12) em uma mistura heterogênea de três subestruturas planas empilhadas, através de ligações π-π, com 3,4 Å de espaçamento e dimensões aproximadas de 10 Å de altura e 20 Å de comprimento.[24–26] Vale destacar que, embora as medidas atuais de microscopia sugerem a formação predominante deste tipo de estrutura, há a formação de outras estruturas poliméricas.[27] É interessante destacar que a presença de espécies polarônicas de nitrogênio, como Ndef[+], não é comumente apresentada na estrutura da melanina. No entanto, existem fortes evidências experimentais da sua presença obtidas por espectroscopia de fotoelétrons excitados por raios X, espalhamento Raman de superfície e espectroscopia infravermelha de que elas existem.[28–33] Tal estrutura está relacionada a subprodutos formados durante o processo síntese.[34]

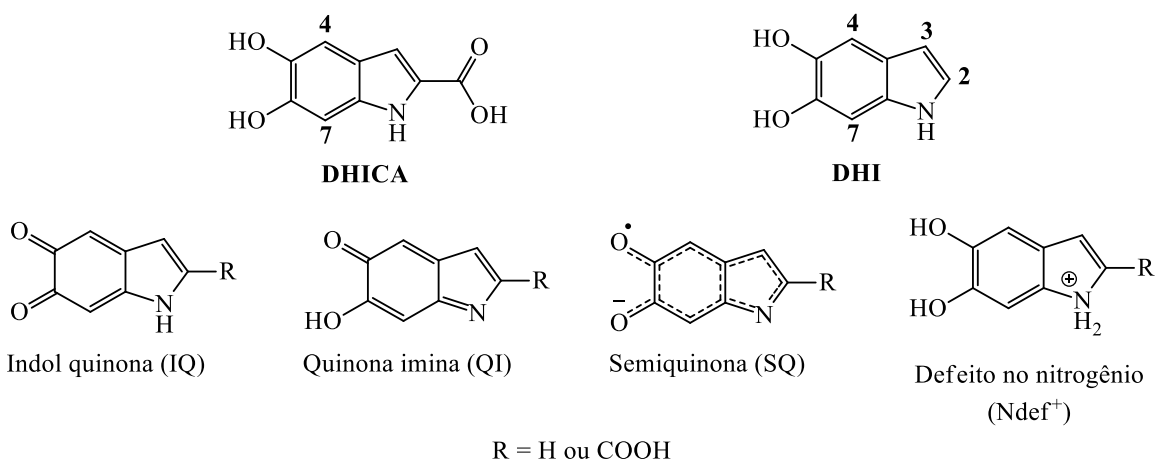

*Figura 1. Monômeros formadores da melanina: ácido 5,6-dihidroxi-indol-2-carboxílico (DHICA) e 5,6-dihidroxi-indol (DHI) em estados redox totalmente reduzidos (denominados por hidroquinona, HQ), IQ totalmente oxidado, QI e SQ parcialmente reduzido/oxidado. Ndef[+] é um subproduto sintético. Para monômeros baseados em DHICA, R = COOH; por outro lado, para DHI, R = H.*

A combinação de estruturas carboxiladas e não-carboxiladas é uma importante característica das melaninas, uma vez que diferentes proporções dos monômeros DHI e DHICA podem influenciar diretamente a estrutura final da macromolécula. Sabe-se que as melaninas ricas em DHICA são formadas por estruturas menores e mais lineares,[2,35] enquanto que as ricas em DHI tem estrutura maiores e com tendência globular. Esta diferença tem relação com o número e com a topologia das posições das ligações cruzadas entre os monômeros. Enquanto DHICA predominantemente polimeriza-se pelas posições C-4 e C-7, DHI pode formar adutos pelas posições C-2, C-3, C-4 e C-7.

# SÍNTESE *IN VIVO* E *IN VITRO*

A biossíntese de melanina, conhecido como processo de melanogênese, ocorre de maneira generalizada em todos os organismos a partir de precursores de tirosina. Nos seres humanos, a melanogênese se dá através da hidroxialquilação da tirosina na presença de oxigênio para formação da L-DOPA e/ou oxidação da L-DOPA para dopaquinona sendo ambos os processos catalisados pela enzima tirosinase (TYR). Entretanto, a dopaquinona é um intermediário bastante reativo que se rearranja e cicliza para gerar o dopacromo.[13,27] O oligômero de melanina final consiste em vários tipos de monômeros que exibem diferentes estados de oxidação, derivados de DHI (Figura 1).[36]

O mecanismo de formação da melanina começou a ser estudado a partir da década de 1920, com os trabalhos de Raper[37] e Mason.[38] Segundo Raper-Mason, o dopacromo seria espontaneamente descarboxilado formando os monômeros DHI, que sofreriam polimerizações oxidativas sucessivas para formação de outros compostos instáveis e finalmente fomação do polímero de melanina.[39] Entretanto, este mecanismo não explica a presença da grande quantidade de grupos carboxílicos, como o presente no DHICA, na melanina natural. Portanto, considera-se que durante a formação da melanina duas enzimas derivadas da tirosinase participam do processo catalítico: proteína-1 relacionada à tirosinase (Tyrp1) e proteína-2 relacionada à tirosinase (Tyrp2), também chamadas de dopacromo tautomerase (DCT). Foi proposto que, na síntese *in vivo*, a enzima Tyrp2 catalisa a tautomerização da dopacromo em DHICA e a Tyrp1 catalisa a oxidação da DHICA, de tal forma a promover sua polimerização,[40] explicando a proporção de DHICA/DHI maior que 50% na melanina natural.[35] Ainda, durante a síntese da melanina, peróxido de hidrogênio ($H_2O_2$) é formado e se acumula no meio reacional.[39–44] O mecanismo de formação do $H_2O_2$ e sua função durante a melanogênese ainda não são completamente compreendidos, mas estudos indicam que tanto $H_2O_2$ quanto o ânion radical superóxido ($O_2^{\cdot-}$) modulam a atividade da tirosinase. [45] Enquanto $H_2O_2$ é um inibidor competitivo da tirosinase, $O_2^{\cdot-}$ ativa essas enzimas 40 vezes. A tironinase também é inibida por tioredoxina e por ditióis. [45] Assim, estas regulações podem ser responsáveis pela alteração na quantidade de monômeros de DHI e DHICA no produto final. Por isso, as condições na qual a síntese ocorre são de extrema importância, visto que pequenas modificações na reação podem afetar a estrutura polimérica e consequentemente as propriedades do material. [35,46,47] Por exemplo, enquanto o processo sintético tradicional de melanina proporciona uma macroestrutura com predominância de DHI, com proporção de DHICA próxima de 10%, na sépia-melanina, um tipo natural de melanina obtida a partir de moluscos marinhos da classe *Cephalopoda*, ocorre uma razão DHICA/DHI maior que 50% (Esquema 1).[35]

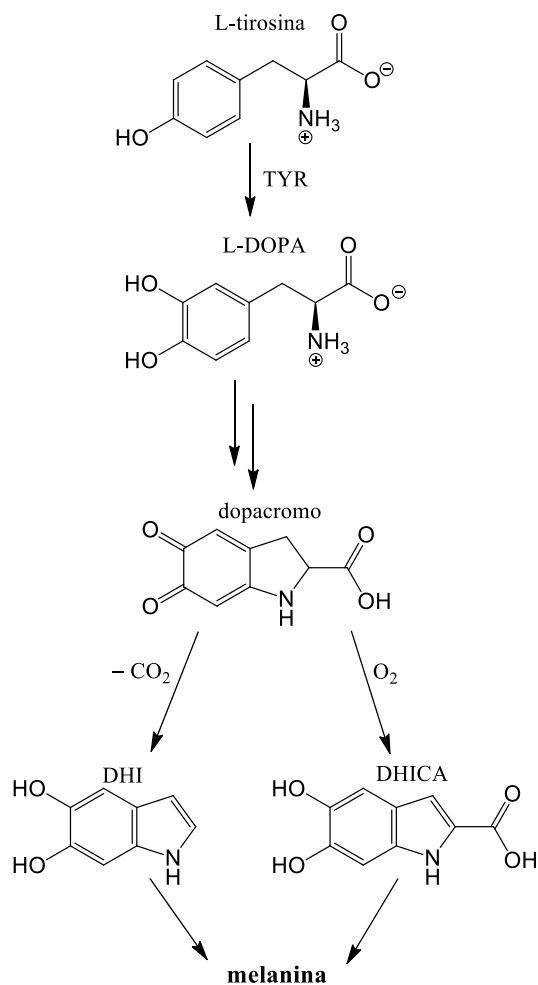

***Esquema 1**. Proposta simplificada da síntese de melanina in vivo.*

Devido a insolubilidade e a dificuldade de extração das melaninas naturais, diversos estudos foram realizados para desenvolver a síntese da melanina *in vitro*. Muitas rotas sintéticas foram propostas para a obtenção de melanina a partir de diferentes precursores, tais como L-DOPA, DHI, DHICA, tirosina e dopamina.[13,48–53] Embora a dopamina venha ganhando bastante atenção devido a facilidade de processamento e a disponibilidade comercial,[54–59] tradicionalmente, a melanina é obtida a partir da oxidação da L-DOPA, por meio de reação enzimática ou pela auto-oxidação em solução aquosa alcalina com o borbulhamento de oxigênio molecular.[13] Ambos os processos sintéticos não são finamente controlados, resultando em materiais poliméricos quimicamente heterogêneos e com diferentes proporções DHICA/DHI. Além disso, assim como na melanina natural, a melanina obtida sinteticamente é insolúvel em água ou em solventes orgânicos, o que dificulta a produção de filmes finos.[60,61]

Bronze-Uhle e colaboradores desenvolveram uma rota sintética mais rápida do que a tradicional também utilizando o borbulhamento de oxigênio molecular. A formação do produto acontece através da polimerização oxidativa da L-DOPA sob pressão de oxigênio em meio alcalino. A nova metodologia resultou em uma melanina sintética homogênea com maior teor de grupos carboxilados, mais mantendo semelhança com a melanina natural. Neste processo, a oxidação da L-DOPA na fase aquosa e meio alcalino é promovida com o $NH_4OH$, para desprotonar as hidroxilas, levando à ciclização e à formação de dopacromo, que segue uma série de etapas oxidativas para formar DHI e DHICA. Nesta rota sintética, a descarboxilação é evitada, o que resultará em uma maior proporção de monômeros DHICA.[31]

A obtenção de um derivado de melanina solúvel foi desenvolvida pelo grupo de pesquisa do Professor Carlos Graeff em 2004.[48,62] Nesta rota, a síntese é controlada o que propicia a obtenção de um material homogêneo, com maior estabilidade térmica e solubilidade em solventes orgânicos como dimetilsulfóxido (DMSO), *N,N*-dimetilformamida (DMF) e *N*-metilpirrolidona (NMP), possibilitando a produção de filmes finos por *spin coating* com melhor aderência nos substratos.[48,62–65]

Analogamente a síntese em meio aquoso, esta rota sintética é baseada na oxidação da L-DOPA, porém, o agente oxidante é o peróxido de benzoíla e o DMSO é utilizado como solvente para substituir a água. Neste ambiente químico, o peróxido de benzoíla além de oxidar a L-DOPA para formação dos monômeros derivados de DHI e DHICA, conforme o modelo de Raper-Mason, também oxida o DMSO para formação do anidrido metanossulfônico.[66] Este subproduto é responsável pela proteção das hidroxilas fenólicas dos monômeros da DHI e DHICA com grupos sulfonados (-$SO_2CH_3$).

Apesar do potencial tecnológico deste material, o tempo sintético é lento, mais de 50 dias. Assim, uma alteração sintética baseada neste mecanismo foi realizada a partir do aquecimento do meio reacional. Observou-se durante o aquecimento da solução à 100 °C que a síntese é ainda mais controlada e o tempo reacional diminui para 8 dias. Além disto, este aquecimento ocasiona uma diminuição dos grupos carboxílicos nos produtos intermediários e, consequentemente, no polímero final.[47] Um resumo do mecanismo proposto para a síntese da melanina em DMSO (DMSO-Melanina) é apresentado no Esquema 2. A partir do mecanismo sintético, foi proposto que a estrutura da DMSO-Melanina é um conjunto de oligômeros com diferentes grupos protetores ligados às hidroxilas fenólicas e ao nitrogênio do anel indol, Esquema 2.

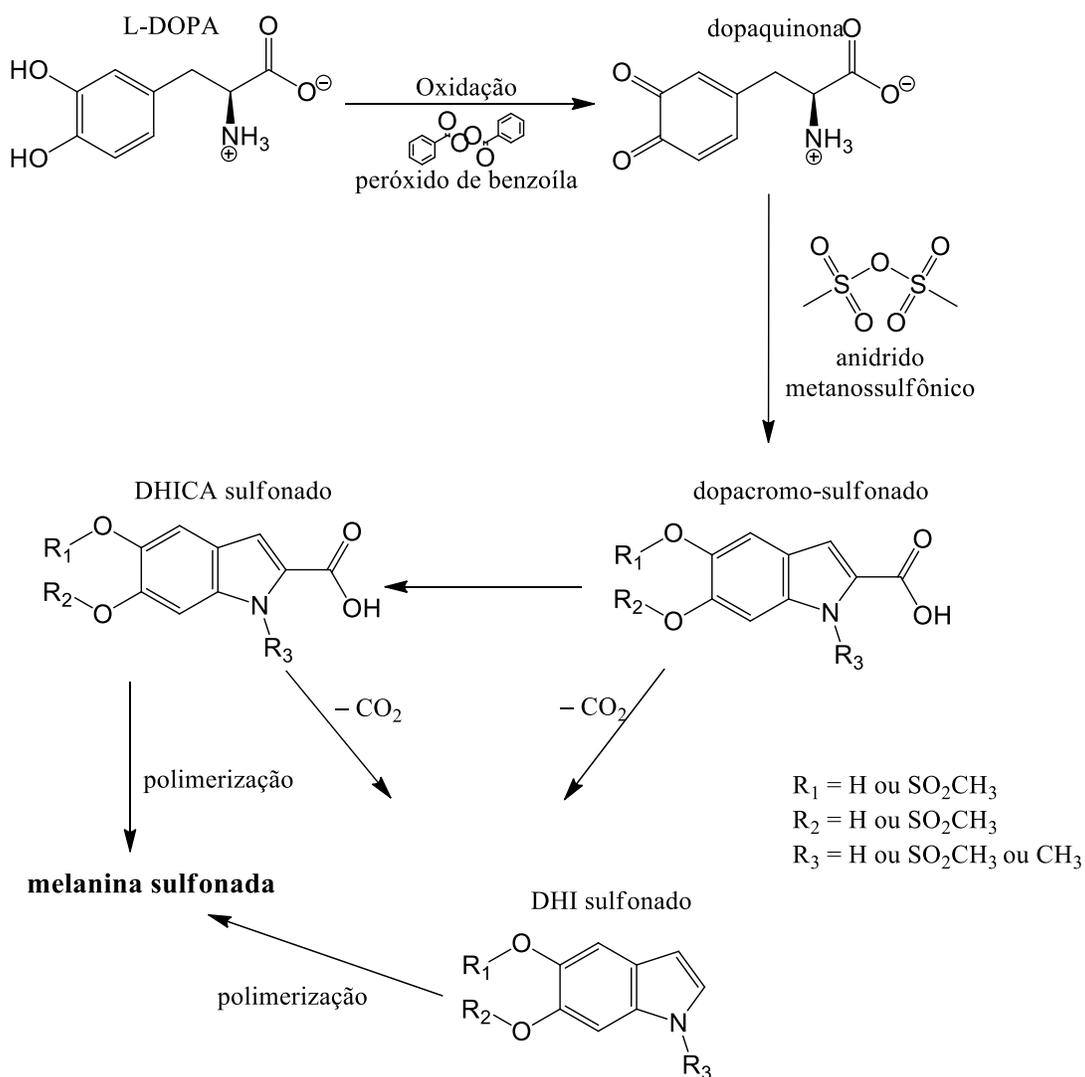

***Esquema 2****. Etapas reacionais propostas para a síntese da melanina sulfonada in vitro.*

A preparação da melanina sintética é um campo com muitos desafios a serem compreendidos e amadurecidos.[67] A melanina é um importante biomaterial e as adaptações em sua rota sintética podem levar a novos produtos multifuncionais com características necessárias para as aplicações desejadas.

## PROPRIEDADES FÍSICO-QUÍMICAS

### Propriedades ópticas

Diferente de muitos pigmentos naturais, a melanina apresenta uma intensa absorção ótica de banda larga, sem nenhum pico proeminente de transição eletrônica, que diminui de intensidade do ultravioleta ao visível e ao infravermelho próximo, o que dá origem a sua coloração escura, comportando-se de maneira equivalente a um semicondutor amorfo.[27]

Atualmente, este comportamento óptico é explicado através de dois modelos de desordens: desordem química e desordem física. No primeiro caso, a absorção óptica decorre da sobreposição dos diversos monômeros que compõe sua estrutura.[68] Este modelo foi reforçado com a demonstração de que os diferentes estados redox contribuem para um espectro amplo de absorção[69] e por mudanças no espectro de emissão com diferentes energias de excitação.[70] Este modelo, contudo, não consegue explicar porque a absorção é maior na região do ultravioleta. Para isto, o modelo da desordem física propõe que tal comportamento seja consequência de folhas de unidades óticas oligoméricas de diferentes tamanhos e empilhadas e aleatoriamente.[71]

### Propriedades paramagnéticas

Outra propriedade que chama bastante atenção nas melaninas é a presença de um sinal paramagnético facilmente detectado em medidas de ressonância paramagnética eletrônica (EPR) em onda contínua.[72] Atualmente, acredita-se que o sinal paramagnético possa ser uma combinação de radicais intrínsecos e extrínsecos. Os radicais intrínsecos, recentemente denominados radicais centrados em carbono (CCR, do inglês *carbono-centered radicals*),[73] são formados durante o crescimento dos oligômeros e agregados da melanina; enquanto os radicais extrínsecos, denominados radicais livres de semiquinona (SFR, do inglês *semiquinone-free radicals*),[73] são associados aos diferentes estados redox dos monômeros da melanina.[72,74–76] Contudo, mesmo após décadas de estudo, ainda não há consenso a respeito da origem estrutural deste sinal, principalmente com relação ao CCR.

Considerando que o sinal de EPR pode ser modificado por uma série de fatores como agentes oxidantes e redutores,[77,78] luz,[79] pH,[80,81] temperatura,[82,83] íons metálicos[84–86] e hidratação[73,87] e seguindo o equilíbrio de comproporcionamento (Esquema 3), acreditava-se que os radicais do tipo semiquinona dessem origem a este sinal. Recentemente, entretanto, através de cálculos eletrônicos da teoria do funcional de densidade, propôs-se que o monômero aniônico do tipo indolquinona também apresenta características espectrais semelhantes ao da semiquinona.[88] Este mesmo trabalho, sugere que as espécies CCR são decorrentes dos monômeros de hidroquinona e de defeitos no nitrogênio neutros ($Ndef^0$).[88] Simulações espectrais dos sinais experimentais reforçaram tais atribuições.[28,87,89]

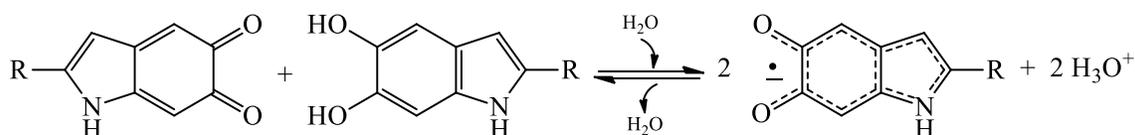

***Esquema 3****. A reação de equilíbrio de comproporcionamento. Espécies completamente oxidadas e reduzidas interagem entre si na presença de água para formarem um intermediário redox (semiquinona) e prótons (na forma de hidrônio).*

**Propriedade de quelação de metais**

A presença de muitos grupos funcionais como amina/nitrogênio indólico, carboxílico, hidroxílico e semiquinona conferem à melanina grande capacidade de ligação com íons metálicos.[26,90] Cátions metálicos como $Ag^+$, $Cu^{2+}$, $Sr^{2+}$, $Cd^{2+}$, $Hg^{2+}$, $Pb^{2+}$, $Fe^{3+}$, $La^{3+}$ e $Gd^{3+}$ possuem grande afinidade para se ligar com a melanina, enquanto que uma menor afenidade de interação ocorre com $Na^+$, $K^+$, $Mg^{2+}$, $Ca^{2+}$ e $Zn^{2+}$.[91–95] O tipo de melanina e o pH podem determinar a eficiência de quelação e quais serão os grupos químicos da melanina que se ligarão aos íons metálicos. [26,90]

**Propriedades elétricas**

No começo dos anos 60 mostrou-se que a deslocalização eletrônica da estrutura indólica da melanina poderia favorecer o transporte de elétrons em modo de semicondutor.[96,97] No entanto, foi só com os dois trabalhos pioneiros de McGinness que o potencial eletrônico das melaninas começou a despertar grande interesse.[98,99] McGinness e colaboradores, mostraram que uma amostra de melanina úmida entre dois eletrodos de ouro apresenta um comportamento de *switch* reversível, compatível com o de um semicondutor orgânico amorfo.[98,99]

Entre os diversos trabalhos que se sucederam, um dos resultados mais importantes foi a caracterização do efeito da hidratação nas propriedades elétricas da melanina, uma vez que a condutividade elétrica varia de $10^{-13}$ Scm$^{-1}$, em uma amostra completamente seca, a $10^{-5}$ Scm$^{-1}$, quando a melanina está completamente hidratada. Além disso, um estudo de coulometria com controle de hidratação mostrou que, quando em estado úmido, os portadores de carga da melanina eram 65% prótons e 35% elétrons.[100,99 100,101]

Outro trabalho bastante relevante foi publicado em 2006 quando a natureza puramente eletrônica do transporte de carga da melanina começou a ser posta à prova. Os autores propuseram que o tipo de carga dominante na melanina era de prótons,[102] que modulariam a densidade de elétrons presentes no sistema, semelhante ao efeito do pH na condutividade de polianilina. Este modelo foi, alguns anos mais tarde, aprimorado. Mostert *et al.* propuseram que as propriedades eletrônicas da melanina não eram de origem semicondutora, mas sim que a adsorção de água daria origem a íons (prótons, H$^+$) e elétrons, seguindo o modelo de comproporcionamento (Esquema 3), de maneira similar ao efeito do pH nas propriedades paramagnéticas da melanina.[103,104] Baseado neste modelo, a melanina é um sistema de condutividade mista iônica-eletrônica, seguindo o fenômeno de auto dopagem química induzida por água.[104] Diversos experimentos posteriores fortaleceram este ponto de vista para as propriedades elétricas da melanina.[65,80,83,84,105,106] No entanto, este não é o fim da linha.

Recentemente, o mesmo grupo de pesquisa questionou o modelo de condutividade mista. Através de medidas de ressonância paramagnética eletrônica induzida por luz e espectroscopia de impedância e dielétrica, sugeriu-se que a melanina deva ser um condutor exclusivamente iônico/protônico, isto é, sem componente eletrônica.[81,107] Porém, outro grupo de pesquisa, liderado pela Professora Santato, revisitou o modelo clássico de semicondutividade e forneceu uma nova perspectiva para transporte protônico em biomateriais. Neste caso, o tipo de portador de carga não é determinado pelas propriedades semicondutoras, mas sim por seus sistemas estendidos de bandas de energia.[108] Neste novo modelo de semicondutividade, a melanina teria ambos os tipos de portadores de carga (iônico e eletrônico), mas que mudariam de uma condução predominantemente eletrônica para uma condução predominantemente iônica/protônica, dependendo do grau de hidratação.[108] Este novo modelo de condutividade poderia explicar tanto o comportamento de switch observado em amostras secas,[109] quanto o comportamento iônico em amostras úmidas.[107]

**Biocompatibilidade**

Testes *in vitro* de biocompatibilidade mostraram que a melanina não impede a adesão e proliferação de células de fibroblastos gengivais humanos,[110] células-tronco embrionárias[111] e células de Schwann.[112] Interessantemente, o derivado sulfonado de melanina mostrou uma maior viabilidade celular quando comparada à melanina padrão.[110]

Testes *in vivo*, por outro lado, demonstraram que filmes de melanina quando implantados em nervos ciáticos possuem baixa propriedades mecânicas e são capazes de provocar uma resposta inflamatório similar à implantes de silicone.[112] Vale mencionar que estes mesmos filmes foram totalmente reabsorvidos após oito semanas.[112]

# ASPECTOS BIOQUÍMICOS

**Melanogênese**

Como já mencionado acima, melaninas são pigmentos poliméricos naturais encontrados em uma variedade enorme de seres vivos, desde bactérias até animais superiores.[113,114] Com composição heterogênea, é responsável por características visuais como a pigmentação da pele, cor dos cabelos e olhos, mas apresenta outras funções, como atividade termorreguladora em vertebrados inferiores e ação fotoprotetiva em peles de mamíferos.[115,116] As melaninas são sintetizadas em células epiteliais especializadas, denominadas de melanócitos, provenientes dos melanoblastos (células derivadas da crista neural embrionária), que se diferenciam em diversos tipos celulares e dentre elas, as células pigmentarias.[117]

No interior dos melanócitos, localizam-se as organelas responsáveis pela síntese e o armazenamento da melanina, os melanossomos, que produzem o pigmento durante a sua própria maturação. Existem 4 estágios de amadurecimento dos melanossomos: o estágio I, também chamado de pre-melanossomo, a organela possui morfologia esférica e matriz amorfa. No estágio II, passa para um formato elíptico, uma matriz fibrilar organizada, porém ainda sem catalisar a síntese do pigmento. Evoluindo para o estágio III, o melanossomo inicia

o processo de síntese da melanina, que se deposita nas fibras internas. Na fase final, estágio IV, a melanina já se deposita sobre todo o interior do melanossomo, que perde sua atividade enzimática, para ser transportado para as células epiteliais vizinhas (queratinócitos) ou para as fibras capilares.[118–121] Variações na distribuição dos melanossomos e fatores como sua composição, número e tamanho, são responsáveis por diferenças observáveis de pigmentação.[122,123]

Como já mencionado detalhadamente acima, a síntese da melanina tem início a partir do aminoácido tirosina, o qual na presença da enzima *tirosinase* e oxigênio molecular e na ausência de cisteína, sofre uma etapa de hidroxilação, gerando L-DOPA, e subsequente oxidação para DOPAquinona (DQ).[124] A DQ é uma molécula extremamente reativa, que produz outros intermediários como a ciclodopa, o dopacromo o DHI e o DHICA, que através de ligações cruzadas formam oligômeros e posteriormente o polímero.[125] Na presença de cisteína/glutationa, a DQ reage formando majoritariamente a 5-S-cisteinildopa e glutationildopa, que após subsequente etapa de oxidação, formam a feomelanina.[126] As concentrações de cisteína e de glutationa são determinantes na produção de um tipo de melanina em detrimento de outro. Baixas concentrações, levarão preferencialmente à formação de eumelanina, enquanto altas concentrações conduzirão à formação da feomelanina. Como já mencionado, $H_2O_2$ e $O_2^{\bullet-}$ são moduladores da atividade da tirosinase, e portanto, enzimas antioxidantes como a *catalase*, *glutationa peroxidase*, *glutationa reductase* e *superóxido dismutase* são cruciais na regulação deste processo. A proporção de grupos tióis conjugados à DQ é diretamente relacionada ao equilíbrio homeostático intracelular e, portanto, o ciclo das pentoses e a regulação do sistema antioxidante das células também estão intimamente ligados ao processo de síntese das melaninas.[126] O balanço hormonal também assume um papel importante, pois no interior dos melanócitos e queratinócitos ocorre a produção da proteína pro-opiomelanocortina (POMC), precursora de vários compostos ativos, como o hormônio estimulante de alfa melanócitos (α-MSH) e o hormônio adrenocorticotrófico (ACTH). Estes peptídeos, ligam-se ao receptor de melanocortina 1 (MC1-R) presentes na membrana celular dos melanócitos, sendo capazes de regular processos de proliferação dos mesmos e estimular a melanogênese.[127,128]

**Propriedades Fotoquímicas e antioxidantes**

Além destes fatores endógenos, a exposição à radiação solar participa de forma ativa no processo de melanogênese. Graças à sua estrutura abundante em duplas ligações, a melanina absorve tanto radiação UV (200 a 400 nm) quanto luz visível (400 e 700 nm), sendo que a absorção luminosa é máxima na região do UVB (290-320 nm).[129] De fato, a radiação solar estimula a produção de proteínas como a POMC, que induzem o processo hormonal citado acima.[130] A faixa de absorção máxima da melanina coincide com a absorção de outras moléculas importantes presente da pele de humanos, como as bases nitrogenadas pirimidinas constituintes do DNA, que ao absorverem radiação UVB podem formar fotoprodutos carcinogênicos.[131,132] Sendo assim, as melaninas exercem um papel importante de proteção contra a mutagênese induzida principalmente por radiação UVB e por isso são reconhecidas como moléculas fotoprotetoras.[123,131–134] No entanto, a melanina também gera espécies excitadas ao interagir com radiação UVA (320-400nm) e luz visível,[116,129,130] gerando espécies tripletes e oxigênio singlete ($^1O_2$), que podem reagir com biomoléculas, gerando efeitos nocivos e até mesmo mutagênicos.[118,132] Além de gerar diversas espécies reativas por excitação eletrônica, as melaninas também reagem com essas espécies reativas, funcionando com agente antioxidante de sacrifício. $^1O_2$, por exemplo, adiciona-se aos anéis indólicos presentes nas melaninas, formando hidroperóxidos e causando o foto-branqueamento das melaninas.[131,132]

A oxidação é um processo inerente ao metabolismo celular, e consequentemente, a formação de radicais e outros compostos reativos ocorre naturalmente em nosso organismo, podendo ser potencializada ou também desencadeada por fatores externos. Alguns dos compostos gerados são genericamente chamados de espécies reativas de oxigênio (EROs),[131] que incluem compostos radicalares como o radical hidroxila (HO$^{\bullet}$), superóxido ($O_2^{\bullet-}$), e peroxila (ROO$^{\bullet}$) e não radicalares, como o peróxido de hidrogênio ($H_2O_2$). Estas espécies podem reagir com moléculas biologicamente ativas, alterando o regime de homeostasia celular e podendo gerar patologias, que podem acometer diversas regiões do corpo humano (pele, olhos, mucosas). O excesso destes compostos oxidantes é anulado através de compostos antioxidantes, que podem sequestrá-los, transformando-os em compostos menos reativos. A ação de um antioxidante pode ocorrer através de mecanismos físicos e/ou químicos. A melanina atua também desta forma, absorvendo a luz e neutralizando espécies radicalares prejudiciais.[132]

Apesar de a melanina ser capaz de sequestrar radicais livres, ela também é capaz de gerá-los após um processo de fotoexcitação por luz UV ou visível.[135] Em geral, a extensão de suas propriedades antioxidantes ou pró-oxidantes serão governadas, de acordo com o tipo de pigmento produzido. Já se constatou que a feomelanina pode estimular a produção de radicais livres ao passo que a eumelanina possui maior capacidade de consumi-los e portanto, ter um efeito predominantemente protetivo.[132] Sendo assim, quanto maior a razão entre eumelanina e feomelanina, menos pronunciada será a manifestação das espécies reativas e, portanto, maior o efeito antioxidante. Estudos em culturas de células epiteliais variando fenótipos de I a VI (o nível de pigmentação aumenta de forma crescente de I a VI), mostram que células contendo maiores concentrações de eumelanina são mais resistentes aos danos fotoquímicos e apresentam maior taxa de sobrevida pós fotossensibilização com radiação UV, em especial UVB.[136] A capacidade da eumelanina de neutralizar os radicais livres depende do número de grupos passíveis de sofrerem oxidação/redução presentes nas unidades, bem como, de seus potenciais redox, e também da facilidade de acesso destes grupos aos radicais.[137]

Durante o processo de polimerização das unidades constitutivas da melanina, ocorre a formação de intermediários que possuem uma alta taxa de transferência de elétrons e de hidrogênio, o que as tornam moléculas susceptíveis à absorção da radiação visível e UV. Após a irradiação dessas entidades, ocorre a formação de radicais superóxido, e estes podem reagir com os intermediários da rota sintética, podendo por exemplo, reduzir quinonas a semiquinonas, e ao mesmo tempo oxidar hidroquinonas a semiquinonas ou quinonas.[138] Como este processo diminui a concentração de radicais superóxido, por consequência também reduz a concentração de radicais hidroxila, que é reativo e consequentemente mais deletério. Outra rota de proteção se dá pela habilidade da melanina em quelar metais, como íons $Fe^{2+}$, o que suprime a reação de Fenton, principal geradora deste último radical.[138,139] Estas evidências corroboram com a hipótese de a melanina ser uma molécula capaz de reduzir os níveis de radicais livres presentes no ambiente celular.

No entanto, com o envelhecimento celular, a exposição constante de células pigmentárias a um ambiente com elevadas concentrações de oxigênio e constante exposição à radiação solar podem depreciar essa atuação, levando até mesmo a causar o efeito inverso, elevando as espécies oxidantes.[140] Essa mudança de atividade pode estar relacionada às modificações estruturais sofridas em decorrência da fotodegradação de suas unidades.[141] Além disso, um dos fatores de maior influência pode ser atribuído ao acúmulo de lipofuscina, um pigmento associado ao envelhecimento celular, que se deposita no citoplasma de células danificadas e que produz oxigênio singlete por fotossensibilização na região espectral do visível.[142,143] A resistência ao envelhecimento e à perda das propriedades antioxidantes da melanina também está diretamente relacionada à proporção entre eumelanina/feomelanina presente.[132]

Para enriquecer as teorias acerca destes fenômenos, muitos estudos têm utilizado seus análogos sintéticos para avaliação de suas propriedades, como habilidade sequestrante de radicais, paramagnetismo, taxas de troca de íons e estrutura.[137] O estudo por meio de modelos parece vantajoso, pois sua síntese é controlada e costumam conter estruturas de elevada solubilidade. Por exemplo, a DOPA e polidopamina (PDA) são polímeros utilizados como modelos de estudo,[144] assim como DHI e DHICA são extensivamente estudados como antioxidantes, graças às suas ligações O-H fenólicas e N-H indólicas, que são capazes de sequestrar radicais livres.[145] Estes compostos, portanto, auxiliam no estudo do comportamento da melanina observado *in vivo*.

**Melanina e proteção solar**

A luz solar apresenta um espectro contínuo de radiação eletromagnética, que é arbitrariamente dividida em três grandes faixas: UV, visível e infravermelho. A radiação UVC é altamente reativa, mas é completamente absorvida pela camada de ozônio, que também absorve parte da radiação UVB.[146–148] A interação da radiação solar com a pele sofre influência das características absorvedoras de luz e fotoquímicas dos cromóforos presentes na pele, que também afeta a profundidade de penetração da radiação na pele.[132,149,150] Sabe-se que a exposição à luz solar pode causar efeitos agudos, como eritema ou bronzeamento, porém a exposição constante e prolongada pode provocar efeitos crônicos como melasma, fotoenvelhecimento, podendo levar inclusive à formação de tumores. Porém, é importante ressaltar que a radiação solar não traz apenas malefícios. Os raios UVB são os principais responsáveis pelo início das reações intracelulares que levam à formação e posterior ativação da vitamina D, e este composto tem uma infinidade de atividades biológicas, podendo inclusive inibir a proliferação de células tumorais, através do estimulo o aumento do número de células apoptóticas.[149,151,152]

Durante o processo de melanogênese, os dendritos presentes nos melanócitos transferem os grânulos de pigmento para cerca de 30 a 40 queratinócitos,[128] realizando seu depósito ao redor do núcleo destas células. Isso não ocorre por acaso, mas devido à capacidade da melanina em proteger o DNA nuclear da radiação UVB.[123] A eumelanina é particularmente responsável por este fenômeno, graças ao seu baixo rendimento quântico de geração de espécies excitadas, ou seja, a razão molecular de formação de transientes reativas por fótons absorvidos é muito baixa.[152] Isso leva a concluir que a eumelanina dissipa a maior parte da energia que recebe em calor, uma característica de moléculas fotoprotetivas.[27,70] Este fato, aliado ao poder sequestrante de radicais livres, mostra que a melanina assume um papel importante e vital na proteção do organismo contra radiação. Já a feomelanina, se mostra um forte fotossensibilizador, uma vez que é capaz de gerar maiores concentrações de oxigênio singlete após exposição à radiação UVA e a luz visível, do que a eumelanina, sob as mesmas condições.[131,152,153] Por isso é importante ressaltar que o comportamento da melanina pode ser variável, dependendo da razão entre as quantidades de cada pigmento.[152–157]

Segundo Meredith & Sarna (2006),[27] podemos classificar a reações fotoinduzidas da melanina em dois subtipos, os processos anaeróbios e aeróbios. O primeiro, pode ser iniciado por radiação UV-visível, em comprimentos de onda acima de 300 nm. Nestes processos, a radiação irá deslocar o equilíbrio redox existente entre as próprias subunidades formadoras da melanina, que se encontram em estado de equilíbrio em suas formas oxidadas e reduzidas, causando transferência contínua de elétrons, que com a dose extra de energia, poderão sofrer deslocamentos e na presença de outros compostos, podem formar novos radicais. No segundo caso, em presença de oxigênio, a radiação UV-visível, em frequências maiores, pode originar as espécies reativas já mencionadas, como $^1O_2$, o $H_2O_2$ e $O_2^{\bullet-}$, responsáveis pela fotodegradação da melanina.[27,155,157]

Pathak et al (1962)[154] realizaram ensaios histoquímicos, confirmando a presença de melanina recém-formada, assim como presença de atividade recente da enzima *tirosinase* em amostras de células de pele humana irradiadas por radiação UV e luz visível, demonstrando que este estímulo implica em aumento do processo de melanogênese. Estes dados abriram caminho para as propostas de mecanismos de formação e estímulo da melanina após processos de fotossensibilização. Hoje sabemos que a radiação UV e a luz visível aumentam a proliferação e diferenciação das células melanocíticas, ao promover a produção dos hormônios precursores da melanogênese nos queratinócitos, como endotelina 1 (estimulador da α-MSH) e a proteína POMC, e estes por sua vez atuam de forma parácrina nos melanócitos com os quais possuem contato.[155] Ela também atua diretamente nas células pigmentárias ao estimular tanto a POMC, quanto seu receptor, e também promovendo enzimas como a *tirosinase* e *Tyrp1*, dentre outros fatores.[156]

Apesar de ainda existirem controvérsias, estudos recentes vêm mostrando uma forma de participação da melanina nas mutações provocadas pela exposição à radiação UV, e consequentemente no aparecimento de patologias como o câncer. Ao sofrerem excesso de exposição ao sol, todo o balanço redox no ambiente celular (homeostase) sofre uma desregulação, que resulta no aumento de espécies oxidativas, como o ânion radical superóxido, já mencionado anteriormente e o radical óxido nítrico (NO•). A junção destes dois compostos pode dar origem à uma terceira entidade, o peroxinitrito ($ONOO^-$). Este ânion é um potente oxidante, naturalmente produzido em nosso organismo, e capaz de degradar a melanina. Uma vez presentes no citoplasma, os fragmentos resultantes da degradação de eumelanina ou feomelanina, tornam-se mais solúveis e são difundidos até o núcleo celular, onde sofrerão troca de energia e irão formar, em uma série de reações, espécies eletronicamente excitadas, capazes de formarem compostos como os dímeros de ciclobutano de pirimidina (CPDs, do inglês *cyclobutane pyrimidine dimers*). A geração de oxigênio singlete também aumenta substancialmente em melaninas degradas, gerando lesões oxidativas em DNA nuclear.[132,155] Estes compostos são atualmente tidos como causadores de mutagenicidade, e logo, podendo sofrer influência da presença de melanina no seu processo de formação.[157–159]

Diante destas evidências, é possível perceber que o papel fotoprotetor da melanina não exclui o fato de que concomitantemente, este biopolímero seja pivô em diversas outras reações, gerando fotoprodutos, radicais livres, intensificando processos de oxidação, decomposição ou até mesmo de mutações e morte celular.[160] Portanto, independentemente do nível de melanina na pele de um indivíduo, deve-se sempre evitar o excesso de exposição ao sol, mesmo com o uso dos filtros solares, uma vez que os efeitos decorrentes da fotossensibilização da melanina ocorrem também na região espectral do visível.[147,161] A ação antioxidante parece ser uma maneira de proteger os melanócitos de danos induzidos por exposição a luz visível.

**APLICAÇÕES TECNOLÓGICAS**

Tendo em vista as diversas propriedades descritas acima, a melanina como uma classe de materiais foi considerada para uma vasta gama de aplicações tecnológicas.

Em relação a aplicações biomédicas, a liberação controlada de partículas de melanina em ferimentos na pele mostrou-se capaz de reduzir a inflamação do ferimento.[162] Além disso, nanopartículas de melanina funcionalizada com PDA[163] ou ainda o recobrimento de matrizes cerâmicas com melanina[164] relevou o potencial do material para *drug-delivery*. Em camundongos, também apresentou efeitos de proteção do fígado contra o estresse induzido pelo álcool[165] ou ainda, em dietas ricas em gordura, prevenção e controle de hiperlipidemia.[166] Melanina tem sido apontada como um biomarcador para melanomas,[167] e também como um agente terapêutico fototérmico capaz de matar células tumorais por hipertermia quando irradiado com radiação na região do infravermelho próximo.[168–170]

Os diferentes estados de oxidação da melanina permitem que ela seja capaz de ligar com fármacos e metais. Esta característica possibilitou a sua utilização em terapia guiada por imagem,[171,172] agentes de contraste fotoacústico e de ressonância magnética[173] e propriedades antibacterianas.[174] Esta mesma propriedade também pode ser utilizada para purificação de águas poluídas com corantes,[175] metais pesados como urânio[176] ou fármacos (como cloroquina).[177] Devido as suas propriedades ópticas, a melanina também foi utilizada como agentes foto-estabilizadores de lã[178] e plásticos.[179] Além disso, também foi mostrado que ela é capaz de aumentar a proteção contra radiação ultravioleta e estabilidade térmica de polímeros.[180]

Do mesmo modo, a melanina é amplamente utilizada na eletrônica orgânica. Nesta área, uma importante característica é a produção de filmes-finos homogêneos. Devido a sua natureza insolúvel, houve grande esforço para obtenção de filmes com qualidade para desenvolvimento de dispositivos nas últimas duas décadas.[61,62,65,111] Todo o desenvolvimento do processamento de filmes de melanina possibilitou a sua utilização como eletrólito sólido em transistores eletroquímicos orgânicos capaz de transduzir corrente iônica em eletrônica.[181] Ou ainda em baterias de sódio[182] e magnésio,[183] capacitores,[184] supercapacitores operando em eletrólitos aquosos[185,186] ou em estado sólido.[187] Aplicações fotovoltaicas como em células solares sensibilizadas por corantes foram estudas com melaninas naturais[188,189] e melaninas eletrodepositadas.[190] Além disso, a melanina também foi utilizada como camada ativa para aplicações em sensores de pH.[191,192]

**CONCLUSÃO E PERSPECTIVAS FUTURAS**

No presente trabalho apresentamos as melaninas e potenciais aplicações. As melaninas vêm despertando um interesse crescente quanto ao seu papel biológico assim como em aplicações na eletrônica e bioeletrônica. Como consequência há um número crescente de grupos de pesquisa estudando-as. Apesar disso, há vários aspectos de sua estrutura e funcionalidade que ainda não são entendidos. Do ponto de vista das aplicações, um dos grandes desafios, que era a produção de filmes finos, parece estar resolvido. Nesta revisão apresentamos vários exemplos envolvendo a combinação de solvente com a funcionalização do polímero. No entanto, muitas questões permanecem, entre elas: qual é a natureza do transporte de cargas, o transporte é iônico, eletrônico ou uma mistura dos dois? Seria possível dopar a melanina e aumentar sua condutividade? Qual a relação do sinal de ressonância paramagnética com o transporte de cargas? Outra questão importante para a aplicação em larga escala é um melhor controle e diminuição do tempo de síntese. Do ponto de vista biológico, discutimos os mecanismos de ativação da melanogênese e as respectivas propriedades fotoquímicas e antioxidantes das melaninas. Não restam dúvidas de que as melaninas exercem um papel protetor e anti-oxidante importante, mas que também podem causar danos em tecidos biológicos. Embora a absorção da radiação UV e da luz visível pelas melaninas ofereça proteção contra danos no DNA das células epiteliais, as melaninas, em especial a feomelanina, também geram espécies reativas, incluindo oxigênio singlete, causando danos oxidativos em DNA por fotossensibilização. Discutimos a consequência desta dicotomia nas estratégias atuais de fotoproteção. Há ainda muitas incógnitas nos mecanismos de transformação maligna que levam melanócitos a se transformar em melanoma, mas já há informações suficiente que sugerem que as melaninas e a melanogênese podem ter um papel importante neste processo.

**AGRADECIMENTOS**

**REFERÊNCIAS**